# Occupational Uniqueness and Wage Growth

*Roberto Vestrelli, University of Perugia, Perugia, Italy*

*Andrea Fronzetti Colladon, Roma Tre University, Rome, Italy*

October 20, 2026

**Abstract**

Changes in occupational task content may have different economic implications depending on whether the tasks gained or lost are occupation-specific or already widespread across the economy. Existing measures of occupational task change largely abstract from this distinction, treating occupations as self-contained task bundles rather than accounting for how tasks are distributed across occupations. We develop a framework to measure a distinct dimension of occupational change, one that goes beyond task expertise or task content: task structure. Using this new framework that tracks tasks across occupations and over time, we separate changes in task quantity from changes in occupational uniqueness, defined as the distinctiveness of an occupation's task bundle relative to other occupations. Using a harmonized dataset of 196,563 task statements from the Dictionary of Occupational Titles and O*NET spanning 1977–2018, we measure task diffusion, occupational uniqueness, and distinct dimensions of task addition and removal. Occupations that become less unique are also associated with lower wage growth, and the wage associations of task additions and removals depend on how widely the affected tasks are performed elsewhere. Adding highly diffused tasks is associated with lower wage growth, as is removing occupation-specific tasks. At the economy-wide dimension, however, the pattern differs: the addition of new tasks is positively associated with wage growth, whereas the removal of tasks is negatively associated with it. Taken together, these results show that similar changes in the number of tasks performed by an occupation can reflect economically distinct processes. Incorporating the cross-occupational distribution of tasks, therefore, provides a richer account of occupational change and its relationship to the wage structure.

## 1. Introduction

Numerous studies documented a shift in the types of activities performed by US workers over the past decades, with much of this change attributed to technological change (Atalay et al., 2020; D. Autor et al., 2024; D. Autor & Thompson, 2025). Previous literature, however, has often neglected how this change can make an occupational task bundle more (or less) unique (that is, different from the tasks that workers in other occupations do). In this paper, we show that changes in the task content of occupations have also caused changes in occupational uniqueness. Contrary to the prediction made by the task-based model (Acemoglu & Restrepo, 2018, 2022), we observe that associations between changes in the task content and wage growth within occupations can be (partly) explained by changes in occupational uniqueness, rather than by the mere addition or removal of tasks. We also show why measures used in previous literature to quantify task additions and removals may show no significant association with wage growth, and provide alternative measures for studying task reconfiguration (D. Autor & Thompson, 2025). To study the antecedents and consequences of changes in occupational boundaries, we create a new harmonized dataset of occupational task content spanning four decades and we document a broad decrease in occupational uniqueness: tasks have become increasingly shared across occupations rather than being concentrated within distinct occupations. We also find that occupations whose task content becomes less unique experience lower wage growth.

Recent work advanced our understanding of how the reconfiguration of job tasks within an occupation relates to changes in wages. For example, Autor and Thompson (2025) developed a theoretical framework in which changes in an occupation's wage are caused by changes in the expertise required to perform the tasks that constitute the occupation. In their framework, new tasks entering an occupation, or obsolete tasks exiting it, can modify the level of expertise required for a worker to enter it. These barriers affect labor supply and, consequently, wages (Acemoglu & Restrepo, 2022; D. Autor et al., 2024).

Similarity between occupations, in terms of the skills needed to perform the tasks, may be an important factor in human capital accumulation and skill transferability, as shown by Gathmann and Schönberg (2010), among others. However, this aspect is usually neglected in the literature on the evolving nature of work. Indeed, the same task can span more occupations, making them more (or less) similar in terms of content, activities, and therefore skills. Put in other words, previous research has treated occupations as self-contained silos by measuring changes in occupational task content only within the boundaries of the occupations themselves. This approach, while useful, leaves unmeasured different dimensions of changes in the nature of work. Once the same task is traced across occupations and time, changes that appear equivalent may represent fundamentally different economic processes. For example, a task added to an occupation may constitute genuinely new work, or it may already be performed elsewhere and simply diffuse across different occupations (causing a decline in what we call occupational uniqueness). Similarly, a task removed from an occupation may completely disappear from the economy or remain part of other occupations. These few examples highlight the importance of a framework for conceptualizing and measuring these changes.

This article proposes a novel, empirical approach for doing so. We construct a new dataset using data from the Dictionary of Occupational Titles (DOT) and the Occupational Information Network (ONET), covering 134,897 tasks between 1977 and 2018. The novelty of the dataset is that it can be used to track whether an occupation includes a task, and how widely that task is diffused in the economy (i.e., how many other occupations perform it). To the best of our knowledge, this aspect is not measured in alternative data sources, and it provides one of the main contributions of this research. We use this new data feature to operationalize two novel constructs. Task Diffusion measures how ubiquitous a task in the economy is by capturing the number of different occupations in which that task appears. Occupational Uniqueness is the corresponding occupation-level measure, capturing the extent to which an occupation's task bundle does not overlap with those of other occupations.

In addition to introducing the two novel constructs, we also extend the economic meaning of two other measures already used in the literature: occupation tasks additions and removals (D. Autor & Thompson, 2025). We do so along two dimensions. First, we distinguish between task additions and removals that are specific to an occupation and those that occur at the economy-wide level. An occupation may undergo a reconfiguration of its task composition when tasks are added to or removed from that occupation alone, or when tasks emerge or disappear from the economy as a whole. We therefore distinguish between Global Task Additions (Removals) and Within-Occupation Task Additions (Removals). Second, we consider how task additions and removals affect occupational uniqueness. Adding a task that is widely diffused across occupations may reduce an occupation's uniqueness. We therefore further distinguish between Unique Task Additions (Removals) and Non-Unique Task Additions (Removals).

Our empirical analysis proceeds in several steps. First, we characterize the average degree of task diffusion in the economy by measuring how widely tasks are performed across occupations and comparing their diffusion in 1977 and 2018. We find that average task diffusion increased over the past 40 years, although the increase varied considerably across task types. We classify each task into four categories: manual, cognitive, routine, and interpersonal, following previous literature (Atalay et al., 2020). In 1977, manual tasks were the least diffused and, therefore, the most occupation-specific. By 2018, cognitive tasks had become the least diffused. This shift is consistent with the technological changes that occurred over the period, due to the introduction of new technologies such as computer-based systems and robots, which have led to the automation of many routine activities with a parallel increased specialization of labor toward cognitive-intensive activities (D. H. Autor et al., 2003). In more recent years, cognitive tasks have therefore become the tasks that most strongly differentiate occupations. By contrast, tasks that require interaction with other people (interpersonal tasks) have become more widely diffused, consistent with the growing importance of service-sector and social skills (Deming, 2017).

Second, we move to the occupational level and ask how occupational uniqueness has changed over time. We find that these changes, between 1977 and 2018, are neither concentrated in specific parts of the earnings distribution nor systematically associated with particular types of work, such as manual, cognitive, or interpersonal tasks, suggesting that changes in occupational uniqueness capture a distinct dimension of occupational content not fully reflected in conventional measures of task content or occupational earnings.

Third, we use our novel measures to study changes in the wage structure of occupations. We link our dataset to wage and employment data from the 1980 Census and the 5-year American Community Survey (ACS) for 2014–2018, using data collected by Autor et al. (2024). We find a strong positive association between changes in occupational uniqueness and wage growth, with occupations that become more unique experiencing higher wage growth. This association also holds when we use composition-adjusted wages, suggesting that it is not simply driven by changes in the demographic characteristics of workers within occupations. Next, we examine task additions and removals to reconcile our findings with the broader literature on occupational task content and wage dynamics. We distinguish between within-occupation and global task changes and find that additions and removals are associated with wage growth only when using global measures. In line with the predictions of task-based models (Acemoglu & Restrepo, 2022), global task additions are positively associated with wage growth, whereas global removals are negatively associated with it. However, these associations are smaller in economic magnitude than those observed for occupational uniqueness. When we re-examine task additions and removals through the lens of occupational uniqueness, we find that their relationship with wage growth depends on how widely tasks are performed across occupations. Adding unique tasks is associated with higher wage growth, while removing them is associated with lower wage growth, consistently with the predictions of task-based models. However, we find the opposite relationship for non-unique tasks. Adding tasks that are already performed in other occupations is associated with lower wage growth, whereas removing them is associated with higher wage growth. These associations also hold when we use composition-adjusted wages. Although our analysis remains associational, our findings suggest that what matters for wage dynamics is not simply whether occupations add or remove tasks, but how these changes affect their occupational uniqueness.

The main contribution of this paper is to provide a novel dataset and a new empirical framework that adds to the literature studying the evolving nature of work. We link our empirical findings to the predictions of the task-based model, which analyzes the wage impact of shifts in the task composition of occupations; furthermore, building on existing research, we demonstrate that accounting for a single dimension (specifically, the fact that a task diffuses across multiple occupations) enables us to derive new measures of occupational change that are significantly associated with wage changes. Another central contribution of the paper is highlighting the link between an occupation's uniqueness and changes in wage growth.

The paper proceeds as follows. The next section relates our framework to the literature on occupational task change and task similarity across occupations. We then describe the data and the method used to harmonize task statements across the DOT and O*NET, construct our measures of task diffusion, and link them to Census and ACS wage data. The results section documents the association between occupational task diffusion and wage growth, along with a decomposition by task additions and removals (within-occupation and global). The final section discusses the implications of our findings and concludes.

## 2. Related Literature

Our paper connects different branches of literature. The first studies how the content of occupations changes over time. Atalay et al. (2020) used historical newspaper job advertisements to track changes in US work from 1950 to 2000. They documented a shift from routine cognitive and manual tasks toward nonroutine analytical and interactive activities. They found that a substantial share of this adjustment has occurred within job titles, rather than across them or occupations. Autor et al. (2024), extending the approach introduced by Lin (2011), studied the emergence of new occupational titles in successive US censuses. The focus of their analysis was to identify the main driving forces behind the creation of "new work", finding that technological change is one of those forces. They showed that labor-augmenting innovations encourage the creation of new work, whereas labor-automating innovations slow down its emergence. More recently, researchers tried to explain why changes in the task content of occupations can affect the wage structure. Autor and Thompson (2025) proposed a framework where task additions and removals from an occupational task bundle alter the level of expertise required to perform the job. Expertise creates barriers to entry and, consequently, affects wages by altering the labor supply. With this paper, we contribute to this literature by providing an alternative (and complementary) explanation of how changes in task content can affect wages. Our main point of departure from the literature cited so far is the observation that tasks can span multiple occupations, and therefore changes in task content can make occupations more or less similar to each other. Gathmann & Schönberg (2010) showed that workers are more likely to move between occupations with similar task requirements, consistent with the transferability of task-specific human capital. Similarly, Cortes and Gallipoli (2018) showed that task similarity across occupations explains part of the mobility costs associated with changing occupations. Besides occupational mobility, overlap in the task content of occupations can also have consequences on wages. Occupational overlap can also make workers more substitutable. Stole & Zwiebel (2003) showed theoretically that greater worker substitutability reduces workers' bargaining power vis-à-vis with firms, thereby exerting downward pressure on wages. The second contribution of this paper is to provide empirical evidence on how task overlap can have an effect on wages. We also provide a methodological framework for measuring longitudinal changes in task overlap by analyzing the task content of occupations.

## 3. Data and Measurement

Our empirical analysis combines two sources of data. We measure occupational task content using the 1977 edition of the Dictionary of Occupational Titles (DOT) and version 23.0 of ONET, released in 2018. We obtain employment, earnings, and demographic information from Autor et al. (2024). This dataset was built using the 1980 Census and the 2014–2018 American Community Survey (ACS) samples distributed by IPUMS USA. Because the main analysis carried out by Autor et al (2024) uses a harmonized taxonomy of occupational categories derived by Autor & Dorn (2013), we use the same taxonomy to build our measures of task change[1].

### 3.1 Occupational Task Data

The DOT and ONET describe the activities performed in detailed US occupations. We use the 1977 DOT and version 23.0 of ONET, released in 2018, for two reasons. First, the two data sources provide a long window for observing changes in occupational task content. Second, these vintages correspond to those used by Autor and Thompson (2025) and Autor et al. (2024), facilitating comparisons with their measures of task additions, task removals, and, more generally, changes in occupational content. The two data sources differ substantially in coverage, structure, and writing conventions. Our extraction produces 114,873 DOT task observations associated with 12,374 detailed source occupation codes and 20,024 O*NET task observations associated with 974 source occupation codes. Because task identifiers and exact wording are not directly comparable across the two systems, we construct a harmonized task network that accommodates textual normalization, semantic equivalence, and one-to-many or many-to-one relationships between task statements. Sections 3.3–3.5 describe our empirical framework to build this task-level network in more detail.

Occupational classifications also change over time. We map the detailed occupations in both sources into the occ1990dd_18 classification. The complete crosswalk defines 306 consistent occupations. After applying source-specific task-data restrictions, the DOT covers 303 harmonized occupations, and O*NET covers 305. The balanced task panel contains the 302 occupations observed in both sources; these are also the occupations retained after merging the task and labor-market data.

[1] The taxonomy used in the paper is occ1990dd_18, This occupational taxonomy is composed of 306 harmonized occupations defined initially by Autor & Dorn (2013) and subsequently revised by Autor et al (2024).

### 3.2 Employment, Earnings, and Demographic Data

Wage and employment data are drawn from the dataset compiled by Autor et al. (2024). In this section, we briefly describe the main variables used in our analysis. For an in-depth explanation of how each variable was built from census data, we refer readers directly to the original paper. All the wage and employment variables in the dataset are computed using data from the IPUMS USA 1980 decennial Census 5-percent sample and 5-year ACS 2014-2018. The underlying worker universe includes individuals aged 17–65 who report having worked during the previous year and who are currently employed in the civilian labor force. The observation level in our main analysis is the occupation, whereas data from the IPUMS census and ACS are at the individual level. When aggregating individual observations at the occupation level, observations are weighted by both the IPUMS person weight (a survey weight directly assigned by IPUMS to an individual based on how many people in the population the individual represents) and the amount of labor a worker supplies over the year. This last weight component is computed by comparing usual weekly hours and actual weeks worked, so that workers supplying more hours (than usual) over the year contribute proportionally more to the occupation-level measure.

We use three different measures of wage changes. The first measure we use, Δ Ln Wage, captures the change in an occupation's average log hourly wage between 1980 and 2018. While this provides a direct measure of how wages change within an occupation, these changes may partly reflect changes in the composition of workers employed in that occupation. For example, an occupation may experience wage growth simply because its workforce becomes more educated, older, or changes in terms of its sex or racial and ethnic composition.

To isolate wage changes from changes in workers' composition, the authors have constructed a predicted wage component. In each period, they have estimated separate cross-sectional regressions of log hourly wages for eight sex-by-race/ethnicity groups (male and female workers within the White, Black, Hispanic, and other race/ethnicity categories). The regressions include indicators for completed schooling and quadratic age profiles fully interacted with schooling (also known as Mincerian regressions). Estimating the models separately across demographic groups allows the relationship between wages, education, and age to vary by sex and race/ethnicity, rather than imposing a common wage structure across these groups. For each worker, the regression fitted value is the log hourly wage predicted by a worker's observable characteristics. Predicted values at the individual level are then aggregated at the occupation level using the same labor-supply weights described above. Δ Ln

Predicted Wage is defined as the change from 1980 to 2018 in the predicted log hourly wage. This variable captures the component of occupational wage growth associated with observable worker characteristics (schooling, age, sex, and race/ethnicity). By subtracting Δ Ln Predicted Wage from Δ Ln Wage, they obtain Δ Ln Adjusted Wage, which measures occupational wage growth after controlling for changes in these observable characteristics.

### 3.3 Harmonizing Task Text

Treating every task without an exact textual counterpart as an addition or removal would confound changes in occupational content with changes in language and task organization. To address this problem, we normalize each task statement by converting the text to lowercase, removing punctuation, English stopwords, and applying stemming. After this procedure, we deduplicate task statements that have identical text across occupations and releases. Text normalization reduces the number of distinct task statements from 134,897 to 98,937, a decline of 26.7 percent. While this harmonization procedure identifies task statements that become identical after normalization, it does not capture substantive continuity when the same activity is described using different words or when potential changes due to editorial practice alter how a task statement is reported. We address this problem and describe the procedure in the next two sections.

### 3.4 A Taxonomy of Task Changes

We start by mapping all the tasks into a high-dimensional text embedding space, using the OpenAI text-embedding-3-small model with 1,536 dimensions. We chose this model for both its performance and its alignment with previous work (D. Autor & Thompson, 2025). After computing the embeddings, for each task, we compute the cosine similarity with any other task. We then select the 25 most similar tasks and manually review them to identify recurring patterns. We find four main patterns (graphically reported in Figure 1 on the left) that we classify as follows:

- *Substantive match.* This first pattern occurs when the focal statement and the candidate match describe the same underlying activity, even though differences in wording may produce a low embedding similarity. For example, the cosine similarity between “Counts bananas” and “Counts or weighs and bundles items” is lower than 0.6. A rule based on cosine similarity would therefore treat these statements as weakly related. However, the lower similarity reflects the more occupation-specific wording of the first statement and the inclusion of additional activities in the second, rather than the disappearance of the underlying task activity. We consider such cases as substantive matches.

- *Consolidation*. This pattern occurs when the focal task is "absorbed" into a broader candidate statement. For example, the DOT task "Reads work order to determine tube tolerance specifications required by customer and type of electric inspection coil to use for tube inspecting" is covered by the broader O*NET statement "Reads work order to determine required processes, techniques, materials, and equipment." The specific activity has not necessarily disappeared; the later source records it as part of a more general task.
- *Split*. This pattern occurs when the focal statement combines activities that the other source records separately. For example, the DOT statement "Installs custom equipment to alter appearance of automotive vehicle according to customer specifications, using handtools and power tools" maps to two O*NET tasks: "Read specifications or confer with customers to determine the desired custom modifications for altering the appearance of vehicles" and "Assembles and installs equipment, using hand tools and power tools."
- *Addition or removal*. This last pattern occurs when the focal task has no substantive counterpart among the candidates and cannot be explained by consolidation, splitting or substantive match. For example, the DOT task "Removes test skein from tube and hangs skein on board designated for each furnace and bushing" has no corresponding O*NET task under the preceding criteria.

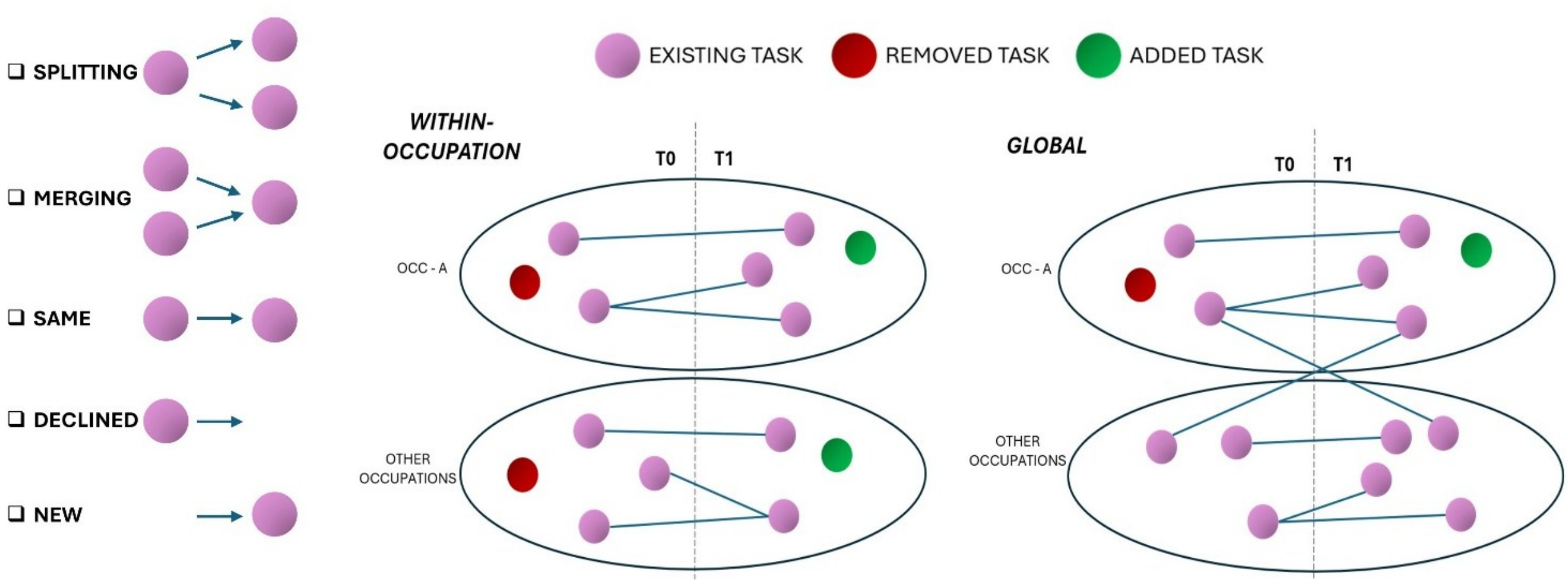

(a)

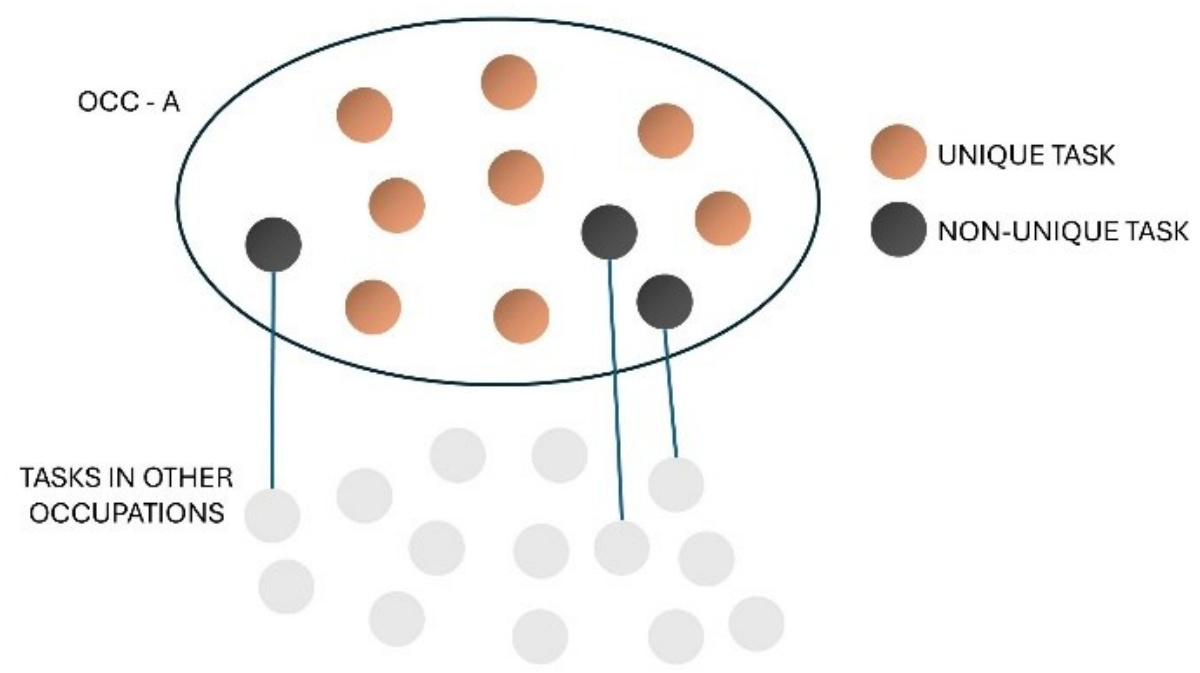

(b)

**Figure 1**. Graphical representation of task connections.

### 3.5 Linking Tasks

A manual review cannot feasibly be extended to every task statement in the panel, considering the size of the dataset. We therefore use GPT-4.1 Mini to apply this taxonomy to all task statements without an exact harmonized counterpart. To do so, we translate the four patterns into two direction-specific classification prompts (we report the full prompts in the appendix). For each task, GPT-4.1 Mini receives the focal task statement in input and a numbered list of its most semantically similar candidate statements (we limit candidate tasks with a cosine similarity of at least 0.50 with a credible assumption that candidates with lower similarity are unrelated in terms of content) and classifies the input task into one of the mutually exclusive categories described. Together with the classification, GPT is also asked to provide links across task statements. Figure 1 on the left provides an intuition for this. For example, if a split occurs, it means the input task can be linked to two or more candidates; in this case, GPT provides the identifiers of the linked tasks (all the prompts are reported in the appendix).

For each focal task, the LLM returns the identifiers of the candidate statements supporting a substantive match, consolidation, or split. We use these identifiers to create links between DOT and O*NET tasks. We use embedding similarity to complement the LLM links. This second procedure captures pairs whose linguistic similarity is exceptionally high relative to the alternative matches available to each task. Rather than applying a common threshold to raw cosine similarity, we standardize similarity separately for every focal task. Let $s_{ij}$ denote the cosine similarity between focal task $i$ and candidate task $j$. We calculate

$$z_{ij}^{(i)} = \frac{s_{ij} - \bar{s}_i}{\sigma_i},$$

where $\bar{s}_i$ and $\sigma_i$ are the mean and standard deviation of task $i$'s similarities with all candidate tasks. We add an embedding link when

$$z_{ij}^{(i)} > 7 \quad \text{or} \quad z_{ij}^{(j)} > 7.$$

The seven-standard-deviation cutoff makes the embedding procedure deliberately conservative: it supplements the LLM classifications only with pairs for which the relative similarity signal is unusually strong[2]. Two tasks with a normalized cosine similarity greater than 7 are likely to be matched with high confidence. The final network is therefore the union of three sets of correspondences: exact harmonized-text matches (1), LLM-supported matches (2), and high-confidence embedding matches (3). LLM links capture relationships that require interpreting the content and scope of the statements. Embedding links match pairs that display exceptional linguistic similarity under the task-specific standardization rule.

For comparisons among tasks within the same source and vintage, we use only links from (1) and (3), without the LLM procedure. This choice is because task statements within the same dataset (all task statements in O*NET or all task statements in the DOT) were produced under common editorial conventions and use more consistent vocabulary. After manually reviewing a subset of 200 classifications within the same dataset, we find that, contrary to comparing statements from different datasets, LLM links from (2) and embedding similarity links from (3) overlap substantially. This reduces the need for (costly) LLM disambiguation. Because of this, to link tasks within the same dataset, we apply the same task-specific standardization and seven-standard-deviation cutoff, linking a pair when the threshold of seven standard deviations is exceeded[3].

---

[2] Our results are robust to alternative cutoffs ranging from four to ten standard deviations. For each cutoff, we reconstruct the task-link network and the resulting occupation-level measures while leaving exact matches and LLM-supported correspondences unchanged. Although the number of embedding links varies mechanically with the cutoff, the main empirical results remain stable.

[3] To build the task network, we also implement an alternative algorithm based on a classification model rather than GPT 4.1. Using the JEV 1.3 model from TypeSafe, we input pairs of sentences into the model and ask it to determine whether they are similar or different. We then construct a network based exclusively on the resulting links. This alternative procedure avoids the explicit classification of tasks into predefined categories and instead relies solely on pairwise links. We also compute all the metrics described in Section 3.6 using this alternative network. Based on these metrics, we obtain results that are both qualitatively and quantitatively similar to those reported in the paper.

### 3.6 Occupational-Level Measures

We use the task links to construct three families of occupational measures: *within-occupation* task additions and removals, *global* task additions and removals, and the *unique (non-unique*) task additions and removals. Let $t \in \{0,1\}$ index the DOT and O*NET vintages; let $o$ denote an occupation; let $i$ index a task. Define $\mathcal{T}_{o,t}$ as the set of distinct tasks observed in occupation $o$ in vintage $t$.

We define a DOT task as removed from an occupation when it has no link to an ONET task in the same occupation. Let $W_i^- = 1$ identify such a task. Symmetrically, $W_i^+ = 1$ when an ONET task has no DOT counterpart in the same occupation. A graphical intuition is provided in the upper-left box of Figure 1. For each occupation, we divide the number of added (removed) tasks by the total number of distinct tasks in the occupation:

$$share_wo_rem_o = \frac{\sum_{i \in \mathcal{T}_{o,0}} W_i^-}{|\mathcal{T}_{o,0}|}$$

$$share_wo_add_o = \frac{\sum_{i \in \mathcal{T}_{o,1}} W_i^+}{|\mathcal{T}_{o,1}|}$$

share_wo_rem and share_wo_add are equivalent to task additions and removals measures defined in Autor et al (2025). These variables capture task reconfiguration within an occupation due to additions (or removals) from the perspective of the occupation itself. This means that a task added to an occupation ($W_i^+$) may already exist at time 0 (the task can be performed by another occupation), equivalently, a task removed from an occupation ($W_i^-$) may still be performed in other occupations at time 1.

Let $GL_i^- = 1$ when a DOT task has no link to any ONET task, regardless of occupation. Let $GL_i^+ = 1$ when an ONET task has no link to any DOT task. The measures of global task removals and additions are defined as follows:

$$share_gl_rem_o = \frac{\sum_{i \in \mathcal{T}_{o,0}} GL_i^-}{|\mathcal{T}_{o,0}|},$$

$$share_gl_add_o = \frac{\sum_{i \in \mathcal{T}_{o,1}} GL_i^+}{|\mathcal{T}_{o,1}|}.$$

The numerator is the number of global tasks added or removed in an occupation task bundle. The denominator is the total number of distinct tasks in the vintage. Compared with the previous set of variables, here a task contributes as an addition (or removal) only if it has no links with tasks in the other dataset (see the right box of Figure 1, Panel a). Thus, the source of occupational novelty (or decline) is assessed by considering the entire task ecosystem rather than changes occurring within a single occupation. As one may note, these measures are, by construction, a subset of the $W$ measures. In other words, a task classified as $GL_i^+$ is also always classified as $W_{i,}^+$, whereas the opposite is not always true.

We next classify within-occupation additions and removals by their diffusion across occupations within the same vintage. For each task $i$ found in occupation $o$, $d_{i,o}$ counts the number of occupations in which $i$ appears. A task is unique when $d_{i,o} = 0$. This happens when the task has no links to other occupations. Conversely, a task is classified as non-unique when $d_{i,o} > 0$. Following this logic, it is possible to define four occupational measures:

$$share_un_rem_o = \frac{\sum_{i \in \mathcal{T}_{o,0}} W_{i,o}^- \ \mathbf{1}\{d_{i,0} = 0\}}{|\mathcal{T}_{o,0}|}$$

$$share_un_add_o = \frac{\sum_{i \in \mathcal{T}_{o,0}} W_{i,o}^+ \ \mathbf{1}\{d_{i,0} = 0\}}{|\mathcal{T}_{o,0}|}$$

$$share_nun_rem_o = \frac{\sum_{i \in \mathcal{T}_{o,0}} W_{i,o}^- \ \mathbf{1}\{d_{i,0} > 0\}}{|\mathcal{T}_{o,1}|}$$

$$share_nun_add_o = \frac{\sum_{i \in \mathcal{T}_{o,0}} W_{i,o}^+ \ \mathbf{1}\{d_{i,0} > 0\}}{|\mathcal{T}_{o,1}|}$$

A task added to an occupation contributes to share_un_add only when the task has zero diffusion within the economy (the task is present only in the occupation of interest). On the other hand, a task added to an occupation contributes to share_nun_add only when the added task is present in (at least) one other occupation in the economy, besides the one of interest. share_un_rem and share_nun_rem follow the same logic. These variables are a decomposition of within-task additions and removals. While in share_wo_add and share_wo_rem any task added (or removed) contributes to the share, here we differentiate based on the diffusion of the tasks added or removed.

We use the task diffusion metric defined above to measure how unique (i.e., differentiated from other occupations) the task content of an occupation is. We first scale the task diffusion of each task $i$ in vintage $t$ by the total number of occupations in that vintage ($|N_t|$). This measure tells us how broadly a task is diffused across other occupations, while taking into account the size of the occupational system in each vintage. We then average task diffusion across the tasks within an occupational task bundle. This gives us our final measure of occupational uniqueness. We multiply the final value by minus one so that the higher the value, the more unique the occupation is, and vice versa.

$$OU_{o,t} = -\frac{\sum_{i \in \mathcal{T}_{o,t}} d_{i,t}}{|N_t||\mathcal{T}_{o,t}|}$$

## 4. Results

### 4.1 Occupational Uniqueness and Wage Growth

In this section, we report three different results. First, occupational boundaries are permeable: tasks appear in several occupations. Second, occupational uniqueness changed substantially between 1977 and 2018, and this change is heterogeneous across occupational groups. Third, occupations that become less unique experience lower wage growth.

Average task diffusion almost doubles over the sample period. It rises from 0.033 percent in 1977 to 0.055 percent in 2018. This difference is significant based on a two-sample t-test ($p < 0.001$). The content of highly diffused tasks also changes. To characterize this change, we classify the 100 most and 100 least diffused tasks in each period into four categories: cognitive and analytical, interpersonal and communicative, administrative and routine, and manual and technical[4]. Table 1 characterizes the composition of the most and least diffused tasks in 1977 and 2018. On each panel, rows are ordered in descending order by the values in the last Column, which reports the relative change between the two vintages. Panel A considers the 100 most diffused tasks. In 1977, almost half of these tasks are labeled as manual and technical. On the other hand, by 2018, interpersonal and communicative tasks constitute the largest category and show the largest increase. The overall picture is that, between 1977 and 2018, the nature of the most diffused tasks shifted from manual and technical to interpersonal and communicative. Panel B considers the 100 least diffused tasks. These tasks are occupation-specific, as they appear in only one occupation across all occupations in each vintage. In 1977, more than half of these tasks are labeled as manual and technical, followed by cognitive and analytical tasks.

[4] We used GPT-4.1 mini to classify each task into one of the predefined categories. The LLM received as input the task statement together with the labels of the available categories and was asked the to select the category that best matched the task.

As in Panel A, Panel B reports the largest decline in manual and technical activities. This decline is consistent with the idea that technologies may have largely automated manual activities, independently of how diffused these tasks were. Compared to Panel A, we find a different pattern for the largest increase, which is displayed by cognitive and analytical tasks. This suggests that what distinguishes occupations from one another lies in the cognitive and analytical activities performed, as also confirmed by the fact that more than one-third of the occupation-specific tasks in 2018 fall within this category.

| **Category** | **DOT** | **O*NET** | **%Δ** |
|---|---|---|---|
| ***Panel A. 100 Most Diffused Tasks*** | | | |
| Interpersonal and communicative | 22% | 39% | +77.3% |
| Administrative and routine | 8% | 13% | +62.5% |
| Cognitive and analytical | 21% | 26% | +23.8% |
| Manual and technical | 49% | 22% | -55.1% |
| ***Panel B. 100 Least Diffused Tasks*** | | | |
| Cognitive and analytical | 19% | 37% | +94.7% |
| Interpersonal and communicative | 14% | 22% | +57.1% |
| Administrative and routine | 9% | 8% | -11.1% |
| Manual and technical | 58% | 33% | -43.1% |

**Table 1**. Distribution of Task Categories by Diffusion in DOT (1977) and O*NET (2018)

In Table 2 we report levels and changes in occupational uniqueness for 12 major occupational groups in 1977 and 2018. The 12 major occupational groups follow the classification as in D. Autor et al. (2024) and are reported on the first Column. The second and third columns report the average occupational uniqueness values within each major group in 1977 and 2018, respectively. Values for occupational uniqueness are standardized to have a mean of zero and a standard deviation of one. The last Column reports the changes between 1977 and 2018, with statistical significance assessed using Welch's two-sample t-test for independent samples.

| **Broad occupational category** | **Occ. Uniqueness 1977** | **Occ. Uniqueness 2018** | **Δ** |
|---|---|---|---|
| Construction and mechanics | -0.498 | 0.227 | 0.725*** |
| Sales minus financial/advertising sales | -0.313 | 0.170 | 0.483*** |
| Production and operative | -0.187 | 0.228 | 0.414*** |
| Farm and mining | -0.153 | 0.229 | 0.382*** |
| Clerical and administrative support | -0.152 | 0.215 | 0.367*** |
| Transportation | -0.034 | 0.225 | 0.259*** |
| Services: Cleaning and protective | 0.057 | 0.201 | 0.144 |
| Managers and executives | 0.243 | 0.232 | -0.010 |
| Service: Personal | 0.188 | 0.138 | -0.050** |
| Technicians + fire, and police | 0.248 | 0.193 | -0.055*** |
| Services: Health | 0.247 | 0.171 | -0.076*** |
| Professionals and financial/advertising sales | 0.242 | -0.673 | -0.915*** |

**Table 2**. Shares and Variation of Occupational Uniqueness by Broad Occupational Category.

Changes in occupational uniqueness vary substantially across the 12 major occupational groups, both in magnitude and direction. The largest decline is observed among Professionals and financial/advertising sales, while smaller decreases are recorded in Service and Technician occupations. No significant changes are observed among Managers and executives or Cleaning and protective services. The largest increases are concentrated in occupational groups located in the middle of the earnings distribution, particularly Construction and mechanics, Transportation, and Clerical and administrative support. However, significant increases are also observed in Farm and mining occupations, which are located at the bottom of the earnings distribution, and in Sales occupations, which are positioned toward the top. These patterns suggest that changes in occupational uniqueness are heterogeneous across the earnings distribution. This heterogeneity also extends to the nature of work performed: occupational groups experiencing the largest increases or decreases do not appear to share a common task profile, whether characterized by manual, cognitive, or interpersonal task intensity. This lack of correlation with work content or earnings suggests that our measure of occupational uniqueness captures something beyond traditional measures of work evolution.

Next, we examine whether changes in occupational uniqueness are associated with changes in wages. For occupation $o$, we estimate this long-difference specification

$$\Delta lnY_0 = \alpha + \beta\,\Delta\mathrm{OU}_o + \gamma_{g(o)} + \varepsilon_o,$$

where the dependent variable is either the difference in log wages (Δ Ln Wage), the difference in composition-adjusted log wages (Δ Ln Adjusted Wage), or the difference in the predicted component of wage (Δ Ln Predicted Wage) between 1980 and 2018. $\Delta OU_o$ denotes the change in occupational uniqueness. To facilitate the interpretation of the coefficients, we express $\Delta OU_o$ in standard deviation units, relative to the 1977 distribution. Thus, an increase of one in $\Delta OU_o$ corresponds to a one standard deviation increase in occupational uniqueness relative to 1980. The term $\gamma_{g(o)}$ denotes fixed effects for the 12 major occupational groups. We weight each observation by the occupation's initial employment share. Table 3 reports the main results.

| | **Δ Ln Wage** | | **Δ Ln Adjusted Wage** | | **Δ Ln Predicted Wage** | |
|---|---|---|---|---|---|---|
| | (1) | (2) | (3) | (4) | (5) | (6) |
| Δ OU | 0.071 | 0.257* | 0.280*** | 0.372*** | -0.211* | -0.117 |
| | (0.151) | (0.152) | (0.102) | (0.096) | (0.109) | (0.095) |
| Occupation Group FE | No | Yes | No | Yes | No | Yes |
| Adjusted $R^2$ | 0.002 | 0.441 | 0.102 | 0.448 | 0.040 | 0.490 |
| Obs. | 302 | 302 | 302 | 302 | 302 | 302 |

*Note*. Robust standard errors are reported in parentheses. All regressions are weighted by employment hours in 1980. Columns (2), (4) and (6) additionally include occupation-group fixed effects. * $p < .05$. ** $p < .01$. *** $p < .001$.

**Table 3**. Regression Estimates of Wage Changes in Response to Changes in Occupational Uniqueness.

We find a positive association between changes in occupational uniqueness and both observed and composition-adjusted wage growth. Although the coefficient in Column (1) is not statistically significant, the inclusion of broad occupational-group fixed effects in Column (2) yields a significant estimate: a one standard deviation increase in occupational uniqueness is associated with a 29.3% increase in wages. This positive association is stronger when considering composition-adjusted wage growth, as reported in Columns (3) and (4). In our preferred specification, Column (4), a one standard deviation increase in occupational uniqueness is associated with a 44.5% increase in composition-adjusted wages. Moreover, changes in occupational uniqueness alone account for more than 10% of the variation in composition-adjusted wage growth. Columns (5) and (6) report negative coefficients for the predicted component of wage growth. This pattern suggest that occupations becoming more unique experience changes in workforce composition toward worker profiles associated with lower predicted wages. However, the association is no longer significant after we include broad occupational group fixed effects. Taken together, the results in Table 3 point to a positive association between OU and changes in wages within occupations. Importantly, this association becomes stronger when we hold constant the demographic profile of workers within occupations, suggesting that the recomposition of workers is not the driver of this effect.

### 4.2 Task Change and Occupational Boundaries

The preceding section documents a broad decline in occupational uniqueness together with a negative association with wage growth. We now ask how changes in occupational task content contribute to that pattern. Two main results emerge. First, wage growth is more closely associated with global task additions and removals rather than with within-occupation changes. Second, the wage association with task reconfiguration measures depends on whether the added (or removed) task is diffuse across occupations or unique to a single occupation.

We estimate the following model:

$$\begin{aligned} \Delta lnY_0 = \; & \alpha + \beta_{WO}^{old} share_wo_rem_o + \beta_{WO}^{new} share_wo_add_o \\ & + \beta_{GL}^{old} share_gl_rem_o + \beta_{GL}^{new} share_gl_add_o + \gamma_{g(o)} + \varepsilon_o, \end{aligned}$$

where, as in the previous specification, the dependent variable can be either the difference in log wages (Δ Ln Wage), the difference in composition-adjusted log wages (Δ Ln Adjusted Wage), or the difference in the predicted component of wages (Δ Ln Predicted Wage) between 1980 and 2018. $\gamma_{g(o)}$ denotes fixed effects for the 12 broad occupational groups and all specifications are weighted by initial occupational employment. The two groups of explanatory variables of interest are denoted by the prefixes wo and gl, respectively. Each group contains two variables measuring task additions and removals, both relative to the total number of tasks in an occupation. Briefly recalling from section 4, the within-occupation (wo) group of variables classifies a task as added (removed) if it has no link within the same occupation in the past (new) vintage. By contrast, the global (gl) group classifies a task as globally added (removed) if the task has no link in any occupation in the previous (new) vintage.

| | (1) | (2) | (3) |
|---|---|---|---|
| *Panel A. Dependent variable: Δ Ln Wage* | | | |
| share_wo_rem | -0.000*<br>(0.000) | | 0.000<br>(0.000) |
| share_wo_add | 0.021<br>(0.020) | | -0.006<br>(0.022) |
| share_gl_rem | | -0.061**<br>(0.019) | -0.062*<br>(0.027) |
| share_gl_add | | 0.109***<br>(0.030) | 0.114**<br>(0.035) |
| Occupation-group FE | Yes | Yes | Yes |
| Adjusted $R^2$ | 0.409 | 0.460 | 0.456 |
| Observations | 302 | 302 | 302 |
| *Panel B. Dependent variable: Δ Ln Adjusted Wage* | | | |
| share_wo_rem | -0.000<br>(0.000) | | -0.000<br>(0.000) |
| share_wo_add | -0.010<br>(0.011) | | -0.020<br>(0.013) |
| share_gl_rem | | -0.030*<br>(0.013) | -0.027<br>(0.016) |
| share_gl_add | | 0.025<br>(0.017) | 0.041<br>(0.022) |
| Occupation-group FE | Yes | Yes | Yes |
| Adjusted $R^2$ | 0.298 | 0.309 | 0.315 |
| Observations | 302 | 302 | 302 |

| | (1) | (2) | (3) |
|---|---|---|---|
| *Panel C. Dependent variable: Δ Ln Predicted Wage* | | | |
| share_wo_rem | -0.000<br>(0.000) | | 0.000<br>(0.000) |
| share_wo_add | 0.031*<br>(0.014) | | 0.013<br>(0.015) |
| share_gl_rem | | -0.031<br>(0.020) | -0.034<br>(0.027) |
| share_gl_add | | 0.083***<br>(0.021) | 0.072**<br>(0.023) |
| Occupation-group FE | Yes | Yes | Yes |
| Adjusted $R^2$ | 0.481 | 0.512 | 0.512 |
| Observations | 302 | 302 | 302 |

*Note*. Robust standard errors are reported in parentheses. All regressions are weighted by employment hours in 1980. * p < .05. ** p < .01. *** p < .001.

**Table 4**. Task Additions, Task Removals, and Occupational Wage Growth.

Panel A of Table 4 reports the results for observed wage growth. Column (1) includes only within-occupation measures of task additions and removals. The coefficient on task removals is negative and significant, although very small in magnitude, while the coefficient on task additions is positive but not significant. Column (2) considers global measures, showing that global task removals are negatively associated with wage growth, whereas global task additions exhibit a positive association. Both relationships with global task additions and removals remain similar in Column (3), where we include both global and within occupation variables in the same specification, while within-occupation task additions and removals lose significance. Global task additions and removals also account for a larger share of the variation in observed wage growth, with an adjusted $R^2$ of 0.460 in Column (2), compared with 0.409 for whitin-occupation task additions and removals in Column (1). Including both sets of variables does not further improve model fit. In terms of economic magnitude, the estimated associations are smaller than those observed in the previous analysis of occupational uniqueness. A 10-percentage-point increase in the share of global additions is associated with approximately a 1.14-percentage-point higher observed wage growth, whereas an equivalent increase in the share of global removals is associated with a 0.62-percentage-point reduction.

Panels B and C examine composition-adjusted wage growth and the predicted component of wage growth, respectively. Overall, the results do not provide conclusive evidence as to whether the associations observed in Panel A primarily reflect changes in workforce composition or in wages, holding worker characteristics constant. In Panel B, Column (2), global task removals are negatively associated with composition-adjusted wage growth, although the coefficient is no longer significant in the full specification reported in Column (3). In Panel C, the most consistent result concerns global task additions, which remain positively and significantly associated with the predicted component of wage growth across specifications. This suggests that the positive association between global task additions and observed wage growth may partly reflect changes in workforce composition toward worker profiles associated with higher predicted wages, rather than wage growth independent of changes in worker characteristics.

To avoid a mechanical correlation between global and within-occupation variables, we also conduct a robustness check by replicating all specifications reported in Table 4 using within-occupation variables that exclude global task additions and removals. The results remain consistent with those reported in Table 4 and are available upon request.

While the WO-GL distinction captures whether a task is added or removed from an occupation or across the economy, it does not describe the task's position relative to other occupations. Global novelty and occupational uniqueness are distinct dimensions. A global task addition may appear simultaneously in several occupations and therefore be non-unique (to the occupation of interest), while an occupation may acquire a task that already exists elsewhere. We therefore further differentiate within-occupation task additions and removals into unique (UN) and non-unique (NUN) tasks. Adding a unique task to an occupational task bundle makes the occupation more unique, while adding a non-unique task has the opposite effect. Conversely, removing a unique task decreases occupational uniqueness, whereas removing a non-unique task increases it, holding constant any other changes in task reconfiguration. We estimate the following model:

$$\begin{aligned} \Delta lnY_0 = \; & \alpha + \beta_{UN}^{old} share_un_rem_o + \beta_{NUN}^{old} share_nun_rem_o \\ & + \beta_{UN}^{new} share_un_add_o + \beta_{NUN}^{new} share_nun_add_o + \gamma_{g(o)} + \varepsilon_o, \end{aligned}$$

where $\Delta lnY_0$ denotes either log change in observed, composition-adjusted or predicted wage in occupation $o$, and $\gamma_{g(o)}$ denotes broad occupational-group fixed effects. We report results in Table 5.

| | (1) | (2) | (3) |
|---|---|---|---|
| *Panel A. Dependent variable: Δ Ln Wage* | | | |
| share_un_add | 5.175**<br>(1.966) | | 3.656<br>(1.995) |
| share_un_rem | -3.509*<br>(1.371) | | -5.022*<br>(2.027) |
| share_nun_add | | -7.344*<br>(2.883) | -4.944<br>(2.905) |
| share_nun_rem | | 1.429<br>(1.366) | -2.225<br>(2.064) |
| Occupation-group controls | Yes | Yes | Yes |
| Adjusted $R^2$ | 0.437 | 0.425 | 0.452 |
| Observations | 302 | 302 | 302 |
| *Panel B. Dependent variable: Δ Ln Adjusted Wage* | | | |
| share_un_add | 4.213*<br>(1.791) | | 1.017<br>(0.961) |
| share_un_rem | -2.911**<br>(0.877) | | -3.232***<br>(0.873) |
| share_nun_add | | -10.800***<br>(1.835) | -10.186***<br>(1.785) |
| share_nun_rem | | 2.505*<br>(1.003) | 0.082<br>(1.050) |
| Occupation-group controls | Yes | Yes | Yes |
| Adjusted $R^2$ | 0.364 | 0.485 | 0.506 |
| Observations | 302 | 302 | 302 |
| *Panel C. Dependent variable: Δ Ln Predicted Wage* | | | |
| share_un_add | 0.912<br>(1.184) | | 2.595<br>(1.338) |
| share_un_rem | -0.558<br>(1.413) | | -1.753<br>(1.890) |
| share_nun_add | | 3.503<br>(1.806) | 5.259*<br>(2.121) |
| share_nun_rem | | -1.109<br>(1.180) | -2.314<br>(1.949) |

| | (1) | (2) | (3) |
|---|---|---|---|
| Occupation-group controls | Yes | Yes | Yes |
| Adjusted $R^2$ | 0.460 | 0.473 | 0.485 |
| Observations | 302 | 302 | 302 |

**Table 5**. Unique and Non-Unique Task Additions and Removals, and Occupational Wage Growth.

Panel A reports the results for observed wage growth. Column (1) includes only unique task measures, showing a positive association for task additions and a negative association for task removals. Column (2) considers non-unique tasks, with additions negatively associated with wage growth, while removals exhibit a negative but not significant relationship. Column (3) includes both sets of measures, showing that unique task removals retain a negative coefficient. In terms of economic magnitude, a 10-percentage-point increase in the share of unique task removals is associated with approximately 39.5% lower wage growth. The full specification also provides the highest explanatory power, with an adjusted $R^2$ of 0.452.

As in previous analyses, Panels B and C further decompose observed wage growth into changes in wages holding workers' demographic characteristics constant (Panel B) and changes attributable to workforce composition (Panel C). Focusing on the full specification in Column (3), the results suggest that the associations between changes in occupational task content and wage growth are primarily reflected in the composition-adjusted component rather than changes in workforce composition. This is also in line with our previous analysis on occupational uniqueness. In particular, unique task removals and non-unique task additions exhibit negative associations with composition-adjusted wage growth, whereas the corresponding associations with the predicted component are weaker or run in the opposite direction. This distinction is particularly evident for non-unique task additions, whose negative association becomes substantially larger in magnitude after holding workforce composition constant. In terms of economic magnitude, a 10-percentage-point increase in the share of unique task removals is associated with approximately 27.6% lower composition-adjusted wages, while an equivalent increase in non-unique task additions is associated with a 63.9% reduction, holding initial wage levels constant.

## 5. Discussion and Conclusions

A growing literature has documented that the content of occupations has changed substantially over the last decades and has linked changes in occupational task content to changes in wages and employment (Acemoglu & Restrepo, 2022; D. H. Autor et al., 2003). However, previous research has not distinguished among the economic processes that can arise when occupations modify their content, which, as we have demonstrated in this paper, can alter how unique occupations are. Drawing on the theoretical framework of task-based models (Acemoglu & Restrepo, 2022), previous work has focused on how task additions and removals affect occupational wage dynamics. A task added to an occupation may signal increased demand for new forms of work, while the removal of a task may occur when the activity becomes obsolete and less in demand in the market. Both of these effects should reflect on wages. These additions and removals also affect occupational uniqueness, raising the question of whether changes in uniqueness help explain differences in wage growth. To investigate this largely unexplored relationship, we introduce a framework for measuring changes in occupational uniqueness and examine their association with wage dynamics. Our analysis focuses on the wage implications of these changes, leaving the investigation of their underlying determinants for future research.

Using a newly harmonized task-by-occupation dataset constructed from the Dictionary of Occupational Titles and ONET, together with a new empirical approach that locates each task within the broader occupational structure, we examine how occupational task content is reconfigured. Our descriptive evidence indicates that the average diffusion of tasks increased between 1977 and 2018, suggesting that tasks have become increasingly shared across occupations. When we move from a task-level to an occupational-level analysis, we find that changes in occupational uniqueness vary substantially across occupational groups. These changes are neither concentrated in specific parts of the earnings distribution nor systematically associated with particular types of work, such as manual, cognitive, or interpersonal tasks, suggesting that changes in occupational uniqueness capture a distinct dimension of occupational content not fully reflected in conventional measures of task content or occupational earnings. When analyzing how the content of unique and diffused tasks changes between 1977 and 2018, we find that interpersonal and communication activities account for an important part of the increase in task diffusion. Activities involving communication and interaction have increasingly become common components of different occupations. This finding provides a complementary perspective on the growing importance of social skills in the labor market documented by Deming (2017). The author showed that employment and wage growth have increasingly favored occupations requiring social interaction and that the labor-market return to social skills has risen substantially. On the other hand, while manual activities were among the more distinctive components of occupations in 1980, cognitive tasks increasingly accounted for the most occupationally specific activities in 2018. Taken together, these findings point to a labor market characterized by two simultaneous developments: a growing set of interpersonal and communication tasks shared across occupations and an increasingly differentiated set of cognitive activities.

We then examine how changes in occupational uniqueness relate to wage growth over nearly four decades, from 1980 to 2018. We find a strong positive association between changes in occupational uniqueness and earnings, with occupations becoming more unique experiencing higher wage growth. Our estimates indicate that a one-standard-deviation increase in occupational uniqueness is associated with a 44.5% increase in wages. One possible explanation for the observed positive relationship is worker sorting: occupations that become more unique may attract workers with characteristics associated with higher earnings, such as greater experience. However, our results hold when using composition-adjusted wages, which control for observable demographic characteristics of workers within occupations. This suggests that the relationship between occupational uniqueness and wage growth is not simply driven by changes in workforce composition but reflects a distinct dimension of occupational change. One possible interpretation of this result is that greater occupational uniqueness affects the relative scarcity and substitutability of the labor services supplied by an occupational group and, consequently, the bargaining power that workers have with respect to the firms (Stole & Zwiebel, 2003). In addition, when a task becomes unique to a single occupation, workers within that occupation may be the only ones accumulating the task-specific human capital required to perform it, potentially allowing them to command higher wages (Gathmann & Schönberg, 2010). We view this mechanism as one possible explanation for the observed empirical relationship. Future research could explore alternative mechanisms and develop theoretical models to better understand the economic forces underlying these findings.

We then move from occupational uniqueness to a more granular analysis of changes in occupational task content, examining task additions and removals. The reason is twofold. First, we want to connect our findings with the broader literature on occupational task content and wage dynamics (D. Autor & Thompson, 2025). Second, we ask whether, and to what extent, the associations between task additions, removals, and wage growth documented in the literature operate through changes in occupational uniqueness.

Regarding the first point, we find that, consistent with previous research, task additions and removals are associated with wage growth. However, these associations are smaller in magnitude than those observed when distinguishing between unique and non-unique task changes, and it remains unclear whether they are driven by changes in worker composition. However, different from previous work, we observe significant associations only for global additions and removals. Because adding a task to an occupation can either increase or decrease its uniqueness, depending on how widely that task is performed across occupations (and the same applies, in reverse, to task removals), we re-examine task additions and removals through the lens of occupational uniqueness and find a clearer picture. For unique tasks, our findings are consistent with the predictions of task-based models: adding a (unique) task is positively associated with wage growth, while removing one is negatively associated with it. Even more interesting, however, is the relationship observed for non-unique tasks, where we find the opposite pattern. Adding a non-unique task, which reduces occupational uniqueness, is negatively associated with wage growth, while removing one is positively associated with it. These associations persist even after controlling for the demographic composition of workers within occupations. Although our analysis is associational, our findings suggest that what matters most for wage dynamics is not simply whether occupations add or remove tasks, but how these changes affect their uniqueness.

The main contributions of this paper are twofold. First, we provide a novel empirical framework for analyzing the evolving nature of work. Our main departure from the previous literature is the development of a method to track the diffusion of tasks across the economy. We do so by combining different methods from textual analysis together with the increased capabilities of LLMs in classification tasks. Building on this framework, we also adopt a network approach to define a set of measures that capture different dimensions of occupational task reconfiguration and to examine how these dimensions are associated with wage changes over the past decades. Second, our contribution relates to the empirical analysis of the potential channels through which these changes are associated with economic outcomes. Previous research has primarily focused on task quantities, job titles, or task expertise (D. Autor & Thompson, 2025). We complement this literature by adding a further dimension: how occupations are interconnected through the activities their workers perform. This perspective allows us to distinguish changes occurring within an occupation from changes in its position relative to other occupations in the economy.

Our framework also suggests several directions for future research. One, as already discussed, is to identify the mechanisms and determinants underlying different types of task reconfiguration. We conjecture that technological change may be one important source of both global task novelty and task decline. At the same time, technological change may also impact occupational uniqueness. New technologies may standardize work or reorganize activities around common tools and technologies, thereby making occupations more similar in terms of the tasks performed and the skills required. An important avenue for future research would therefore be to identify which types of technological shocks generate global novelty and which instead induce convergence across occupations, thus reducing occupational uniqueness. A second direction is to investigate more directly the mechanisms underlying the negative relationship between declining occupational uniqueness and wage growth. If this relationship operates through greater worker substitutability, future research could examine whether the association is stronger in labor markets or in institutional settings where employers can more easily substitute workers across occupations. More broadly, distinguishing between changes in what workers do and changes in how substitutable their work becomes may provide a useful lens for understanding how the evolving organization of tasks shapes the wage structure.

## References

Acemoglu, D., & Restrepo, P. (2018). The Race between Man and Machine: Implications of Technology for Growth, Factor Shares, and Employment. *American Economic Review*, *108*(6), 1488–1542. https://doi.org/10.1257/aer.20160696

Acemoglu, D., & Restrepo, P. (2022). Tasks, Automation, and the Rise in U.S. Wage Inequality. *Econometrica*, *90*(5), 1973–2016. https://doi.org/10.3982/ECTA19815

Atalay, E., Phongthiengtham, P., Sotelo, S., & Tannenbaum, D. (2020). The Evolution of Work in the United States. *American Economic Journal: Applied Economics*, *12*(2), 1–34. https://doi.org/10.1257/app.20190070

Autor, D., Chin, C., Salomons, A., & Seegmiller, B. (2024). New Frontiers: The Origins and Content of New Work, 1940–2018*. *The Quarterly Journal of Economics*, *139*(3), 1399–1465. https://doi.org/10.1093/qje/qjae008

Autor, D. H., & Dorn, D. (2013). The growth of low-skill service jobs and the polarization of the US labor market. *American Economic Review*, *103*(5), 1553–1597.

Autor, D. H., Levy, F., & Murnane, R. J. (2003). The Skill Content of Recent Technological Change: An Empirical Exploration. *The Quarterly Journal of Economics*, *118*(4), 1279–1333. https://doi.org/10.1162/003355303322552801

Autor, D., & Thompson, N. (2025). Expertise. *Journal of the European Economic Association*, *23*(4), 1203–1271. https://doi.org/10.1093/jeea/jvaf023

Cortes, G. M., & Gallipoli, G. (2018). The Costs of Occupational Mobility: An Aggregate Analysis. *Journal of the European Economic Association*, *16*(2), 275–315.

Deming, D. J. (2017). The growing importance of social skills in the labor market. *The Quarterly Journal of Economics*, *132*(4), 1593–1640.

Gathmann, C., & Schönberg, U. (2010). How General Is Human Capital? A Task-Based Approach. *Journal of Labor Economics*, *28*(1), 1–49. https://doi.org/10.1086/649786

Lin, J. (2011). Technological Adaptation, Cities, and New Work. *The Review of Economics and Statistics*, *93*(2), 554–574. https://doi.org/10.1162/REST_a_00079

Stole, L. A., & Zwiebel, J. (2003). Organizational Design and Technology Choice under Intrafirm Bargaining: Reply. *American Economic Review*, *93*(1), 456–457. https://doi.org/10.1257/000282803321455421

## Appendix

This appendix reports the two prompts used for the LLM-based classification described in Section 3.4. We use separate prompts depending on the direction of the comparison: from DOT to ONET and from ONET to DOT.

The four conceptual patterns described in the main text are operationalized differently depending on the direction of the mapping. Substantive match is classified as SAME in both directions. The addition or removal pattern is classified as DECLINED when a DOT task has no substantive counterpart in ONET and as NEW when an ONET task has no substantive counterpart in DOT. The other two patterns require direction-specific labels because the same restructuring appears differently depending on which task is taken as the focal statement. A consolidation is classified as MERGING when moving from DOT to ONET, because an earlier task may be absorbed into a broader later task, and as CONSOLIDATION in the reverse direction, when a later task combines multiple earlier tasks. Similarly, a split is classified as SPLITTING when a DOT task is distributed across multiple ONET tasks and as UNBUNDLING in the reverse direction, when a later task appears as a more specific component of a broader earlier task. Thus, MERGING and CONSOLIDATION represent the same conceptual pattern viewed from opposite directions, as do SPLITTING and UNBUNDLING.

The complete prompts are reported below.

### *DOT-to-O*NET classification prompt*

You are given as input:

1. A task statement from the {VERSION_0} version of the {ONET_DOT_0_ext}.
2. A short, numbered list of the closest semantically similar task statements from the {VERSION_1} version of the {ONET_DOT_1_ext}, identified by a matching procedure based on cosine similarity of embeddings.

Your role is to classify the {VERSION_0} {ONET_DOT_0} task into exactly one of the following categories.

Classification categories:

1 - SAME
Classify the {VERSION_0} {ONET_DOT_0} task as SAME if at least one listed {VERSION_1} {ONET_DOT_1} task can reasonably be considered the same task as the {VERSION_0} {ONET_DOT_0}, despite differences in wording or phrasing.

Indicate which {VERSION_1} task number(s) appear to correspond to the same task.

2 - DECLINED
Classify the {VERSION_0} {ONET_DOT_0} task as DECLINED if none of the listed {VERSION_1} {ONET_DOT_1} tasks can reasonably be considered the same task, and the absence of a covering {VERSION_1}

{ONET_DOT_1} task can plausibly be attributed to automation, mechanization, computerization, obsolescence, or substantial decline in the task as human-performed work between {VERSION_0} and {VERSION_1}.

Automated means that firms by {VERSION_1} commonly automated this type of repetitive task at production scale. It does not require that every firm, workshop, or establishment automates the task.

Declined as a human-performed task means that the task became substantially less common as work performed directly by humans between {VERSION_0} and {VERSION_1}. This includes cases where the task may still exist but is now often performed by machines, embedded in automated production systems, or no longer appears as a distinct human task.

3 - MERGING
Classify the {VERSION_0} {ONET_DOT_0} task as MERGING if the task is not listed in {VERSION_1} as a separate task, but appears to have been absorbed into one or more broader {VERSION_1} {ONET_DOT_1} tasks among the listed candidates.

Indicate which {VERSION_1} task number or numbers the {VERSION_0} task appears to have been absorbed into.

4 - SPLITTING
Classify the {VERSION_0} {ONET_DOT_0} task as SPLITTING if the {VERSION_0} task appears to have been distributed across multiple listed {VERSION_1} {ONET_DOT_1} tasks, such that no single {VERSION_1} task fully captures it, but several listed {VERSION_1} tasks together cover its main components.

Indicate which {VERSION_1} task numbers jointly cover the {VERSION_0} task.

Important decision guidance:

- Do not classify a task as DECLINED merely because automation is technically possible.
- Do not classify a task as DECLINED merely because the {VERSION_0} {ONET_DOT_0} wording sounds old-fashioned, manual, industrial, repetitive, or highly specific.
- Do not classify a task as still human-performed merely because some humans may still do it in small shops, craft production, repair settings, or exceptional cases.
- The absence of a covering {VERSION_1} {ONET_DOT_1} task among the listed matches is not by itself evidence of automation or decline, because O*NET task statements are representative exemplars rather than exhaustive lists of all tasks performed in an occupation.

Input format:

{VERSION_0} {ONET_DOT_0} task:
<task statement>

Closest {VERSION_1} {ONET_DOT_1} task statements:
1. <task statement>
2. <task statement>
3. <task statement>
...

Output format:

Return only valid JSON:

```
{
  "label": <1|2|3|4|5>,
  "category": "<SAME|DECLINED|MERGING|SPLITTING|MEASUREMENT_NOISE>",
  "matched_onet_task_numbers": [<number>, ...] or null,
  "rationale": "<one or two concise sentences explaining the decision>"
}
```

### *O*NET-to-DOT classification prompt*

You are given as input:

1. A task statement from the {VERSION_1} version of the {ONET_DOT_1_ext}.
2. A short, numbered list of the closest semantically similar task statements from the {VERSION_0} version of the {ONET_DOT_0_ext}, identified by a matching procedure based on cosine similarity of embeddings.

Your role is to classify the {VERSION_1} {ONET_DOT_1} task into exactly one of the following categories.

Classification categories:

1 - SAME
Classify the {VERSION_1} {ONET_DOT_1} task as SAME if at least one listed {VERSION_0} {ONET_DOT_0} task can reasonably be considered the same task as the {VERSION_1} {ONET_DOT_1} task, despite differences in wording or phrasing.

Indicate which {VERSION_0} task number(s) appear to correspond to the same task.

2 - NEW
Classify the {VERSION_1} {ONET_DOT_1} task as NEW if none of the listed {VERSION_0} {ONET_DOT_0} tasks can reasonably be considered the same task, and the absence of a covering {VERSION_0} {ONET_DOT_0} task can plausibly be attributed to the emergence, diffusion, or substantial growth of the task between {VERSION_0} and {VERSION_1}.

3 - UNBUNDLING
Classify the {VERSION_1} {ONET_DOT_1} task as UNBUNDLING if the task appears to have existed in {VERSION_0} as part of one or more broader {VERSION_0} {ONET_DOT_0} tasks, but appears in {VERSION_1} as a more specific or separately listed task.

Indicate which {VERSION_0} task number or numbers the {VERSION_1} task appears to have been unbundled from.

4 - CONSOLIDATION
Classify the {VERSION_1} {ONET_DOT_1} task as CONSOLIDATION if the {VERSION_1} task appears to combine, generalize, or consolidate multiple listed {VERSION_0} {ONET_DOT_0} tasks, such that no single {VERSION_0} task fully captures it, but several listed {VERSION_0} tasks together cover its main components.

Indicate which {VERSION_0} task numbers jointly cover the {VERSION_1} task.

Important decision guidance:

- Do not classify a task as NEW merely because no close {VERSION_0} task is listed.
- Do not classify a task as NEW merely because it uses modern-sounding terminology.
- The absence of a covering {VERSION_0} {ONET_DOT_0} task among the listed matches is not by itself evidence that the task was genuinely new, because O*NET task statements are representative exemplars rather than exhaustive lists of all tasks performed in an occupation.

Input format:

{VERSION_1} {ONET_DOT_1} task:
<task statement>

Closest {VERSION_0} {ONET_DOT_0} task statements:
1. <task statement>
2. <task statement>
3. <task statement>
...

Output format:

Return only valid JSON:

```
{
  "label": <1|2|3|4|5>,
  "category": "<SAME|NEW|UNBUNDLING|CONSOLIDATION|MEASUREMENT_NOISE>",
  "matched_onet_task_numbers": [<number>, ...] or null,
  "rationale": "<one or two concise sentences explaining the decision>"
}
```